\documentclass[prd,twocolumn,showpacs,showkeys,preprintnumbers,floatfix,
  nofootinbib,superscriptaddress]{revtex4-1}
\usepackage{amsfonts} % AMS
\usepackage{amssymb} % AMS
\usepackage{amsmath} % AMS
\usepackage{subfigure} 
\usepackage{graphicx} % Include figure files
\usepackage{array} % array
\usepackage{dcolumn} % Align table columns on decimal point
\usepackage{bm} % bold math
\usepackage{latexsym} % latex symbols
\usepackage{longtable} % long tables
\usepackage{hyperref} % hypertext links 
\begin{document}

%-------------------------
% \preprint{SNUTP-04-008}
%-------------------------

\title{Finite volume effects in $B_K$ with improved staggered fermions}
\author{Jangho Kim}
\affiliation{
  Lattice Gauge Theory Research Center, FPRD, and CTP,\\
  Department of Physics and Astronomy,
  Seoul National University, Seoul, 151-747, South Korea
}
\author{Chulwoo Jung}
%
%\email[E-mail at ]{chulwoo@bnl.gov}
%
% \homepage[Home page at: ]{http://quark.phy.bnl.gov/www/who.html}
%
\affiliation{
  Physics Department, Brookhaven National Laboratory,
  Upton, NY11973, USA
}
\author{Hyung-Jin Kim}
\affiliation{
  Lattice Gauge Theory Research Center, FPRD, and CTP,\\
  Department of Physics and Astronomy,
  Seoul National University, Seoul, 151-747, South Korea
}
\author{Weonjong Lee}
\email[E-mail:]{wlee@snu.ac.kr}
\homepage[Home page:]{http://lgt.snu.ac.kr/}
\altaffiliation[Visiting professor at ]{
  Physics Department,
  University of Washington,
  Seattle, WA 98195-1560, USA
 }
\affiliation{
  Lattice Gauge Theory Research Center, FPRD, and CTP,\\
  Department of Physics and Astronomy,
  Seoul National University, Seoul, 151-747, South Korea
}
\author{Stephen R. Sharpe}
%
%\email[E-mail at ]{sharpe@phys.washington.edu}
%
%\homepage[Home page at ]{http://www.phys.washington.edu/users/sharpe/}
%
\affiliation{
  Physics Department,
  University of Washington,
  Seattle, WA 98195-1560, USA
}
\collaboration{SWME Collaboration}
\date{\today}
\begin{abstract}
We extend our recent unquenched ($N_f=2+1$ flavor) calculation
of $B_K$ using improved staggered fermions by including
in the fits the finite volume shift predicted by one-loop staggered
chiral perturbation theory.
The net result is to lower the result in the continuum limit by
$0.6\%$. This shift is slightly smaller than our 
previous estimate of finite volume effects 
based on a direct comparison between different volumes.
%
%To include the finite volume effects in a reasonable time, 
%we found it necessary to
%calculate them using Graphics Processing Units.
%
\end{abstract}
\maketitle
%
%
%

%\section{Introduction \label{sec:intr}}
%
We have recently reported a result for $B_K$
using improved staggered fermions with all errors 
controlled and with a total error of 6\%~\cite{wlee-10-3}. 
(We refer to this paper as SWME in the following; a recent
update including an additional lattice spacing is given
in Ref.~\cite{wlee-10-7}.)
Although the dominant error in this result is from uncertainty
in the matching factors, an important subdominant source of error
is that arising from our use of a finite volume.
In SWME we estimated this error to be 0.85\% by comparing the
result obtained on two lattices of different volumes 
(ensembles C3 and C3-2, as discussed below).
Here we revisit the finite volume (FV) error
using an alternative approach: repeating our chiral fits
using forms predicted by one-loop SU(2) 
staggered chiral perturbation theory (SChPT)
but now including FV effects from pion loops.

We did not previously include one-loop FV effects because
we found that fitting with the FV form using CPUs 
on our extensive dataset 
%(in our case Intel i7 920's)
took too long (of order two months). 
% to be useful since one needs many analyses to estimate errors.
We now use GPUs (Graphics Processing Units)
%---specifically Nvidia GTX 480's), 
which allows us to
reduce the fitting time down to a few hours, 
while maintaining double-precision throughout.
Thus we can undertake the many analyses needed to estimate
systematic errors.

%This report serves as an addendum to SWME, and
%we do not repeat many details described in that work.
%Thus, in the following, we describe first the changes necessary
%to add FV effects to the SChPT prediction, present the
%results of the new fits, and then conclude.
%A partial and preliminary account of this work appeared in
%Ref.~\cite{wlee-10-6}.

%\section{FV effects in SU(2) staggered chiral perturbation theory\label{sec:su2}}
%
To extrapolate our data to the physical light quark masses and
to the continuum limit we use the functional form predicted by
SChPT\cite{wlee-99,bernard-03}.
Specifically, our central result in SWME used
partial next-to-next-to-leading order (NNLO)
SU(2) SChPT fits, and we consider here only such fits.
The key extrapolation is in the valence down-quark mass,
$m_x$, or more precisely in the mass-squared of the valence pion
with flavor $\bar x x$, which we call $X_P$.
The NNLO form contains 3 parameters (at a fixed lattice spacing $a$ and
fixed sea-quark masses):
the coefficients of (i) the constant $+$ chiral logarithm term,
(ii) the term linear in $X_P$ 
and (iii) the term quadratic in $X_P$.
For details see Eqs.~(41-45) of SWME.
Finite volume corrections only impact the chiral logarithms.
In SWME these are written in terms of the functions:
\begin{eqnarray}
\ell(X) &=& X \left[\log(X/\mu_{\rm DR}^2) +\delta^{\rm FV}_1(X) \right]\,,
\label{eq:app:l}
\\
\tilde\ell(X) &=& - \frac{d\ell(X)}{dX} 
=
-\log(X/\mu_{\rm DR}^2) -1 +\delta^{\rm FV}_3(X)\,,
\label{eq:app:tilde-l}
\end{eqnarray}
where $\mu_\text{DR}$ is the scale introduced by dimensional
regularization.
%(which, as discussed in SWME, we set to
%$\mu_\text{DR}=0.77\;$GeV).
%
The FV parts, left out in SWME but included here, are
\begin{eqnarray}
\delta^{\rm FV}_1(M^2) &=& \frac4{ML} \sum_{n \ne 0}
\frac{K_1(|n| ML)}{|n|}
\label{eq:delta_1}
\\
\delta^{\rm FV}_3(M^2) &=& 2 \sum_{n\ne 0}
{K_0(|n| ML)}\,,
\label{eq:delta_3}
\end{eqnarray}
where $L$ is the spatial box size
and $n=(n_1,n_2,n_3,n_4)$ is a vector of integers labeling image
positions.
The norm $|n|$ is defined as
\begin{equation}
|n| \equiv \sqrt{n_1^2 + n_2^2 + n_3^2 + \left(\frac{L_T}{L} n_4\right)^2}
\,,
\end{equation}
with $L_T$ is the (Euclidean) temporal box size.
$K_0$ and $K_1$ are the standard modified Bessel
functions of the second kind, which fall exponentially for large $x$.

The expressions (\ref{eq:delta_1}) and (\ref{eq:delta_3}) for the 
FV corrections cease to be valid when $ML\lesssim 1$. One then moves from
the so-called ``p-regime'' into the ``$\epsilon$-regime'', and
a different power-counting applies. Our calculations are done with
valence pions satisfying $ML \gtrsim 3$, which, as the following results
indicate, appears to be above the minimum value at which
it is appropriate to use the above expressions.

%\section{Numerical results \label{sec:fit-su2}}

%\subsection{Lattice details}
%
The lattice ensembles used in this paper are those
used in SWME to do the continuum extrapolation,
namely (in the notation of SWME) C3 (coarse),
F1 (fine), S1 (superfine), together with a new 
ensemble U1 (ultrafine). The latter has a nominal lattice spacing of
$0.045\;$fm, sea quark masses $am_\ell=0.0028$ and $am_s=0.014$,
and size $64^3\times 192$. Our results for this fourth lattice
spacing are preliminary, based on only 305 configurations.
We also use the larger volume coarse ensemble, C3-2.

On each ensemble we calculate $B_K$ using 10 different sea
quark masses, ranging down from $\approx m_s^{\rm phys}$ to
$\approx m_s^{\rm phys}/10$. The precise values are given
in SWME for the S1, F1, C3 and C3-2 ensembles, while for the U1
ensemble they are $0.0014 \times n$ with $n=1,2,3,\ldots,10$.
For the chiral extrapolation we use the lightest four valence
masses for $m_x$, while for the valence strange quark (which we call $m_y$)
we use the three heaviest valence masses to extrapolate to the
physical strange-quark mass.
%Thus we keep $m_x \ll m_y \sim m_s^\text{phys}$,
%as is necessary in order to use SU(2) SChPT.
We do the fit in two stages, first (the ``X fit'')
extrapolating in the valence
$d$ quark mass to the physical value while holding the valence $s$ quark
mass fixed, and second (the ``Y fit'') extrapolating linearly to the
physical valence $s$ quark mass.
Altogether this procedure is labeled a ``4X3Y'' fit.

%Details of the lattice operators used to calculate $B_K$, the determination
%of fitting ranges in Euclidean time $T$, and the fitting function of $T$,
%are presented in SWME-1 and not repeated here.
%
%Details of the one-loop matching factors are given in
%Refs.~\cite{wlee-10-4,wlee-10-1}.
%
%Our calculation relies on the assumption
%that using SChPT takes care of the unphysical effects 
%introduced by using a rooted determinant,
%following the arguments given in 
%Refs.~\cite{Bernard:2006zw,Sharpe:2006re,Bernard:2007ma,Shamir:2006nj}.

%All fits use only the diagonal part of the (inverse) correlation matrix,
%since we have not been able to do stable fits with the full
%correlation matrix. For further discussion of this point see Refs SWME-1
%and \cite{wlee-11-1}.

%\subsection{Fits including FV effects}
%
%
To calculate the finite volume corrections 
$\delta^{\rm  FV}_1$ and $\delta^{\rm FV}_3$, 
we must truncate the sums in
Eqs.~(\ref{eq:delta_1}) and (\ref{eq:delta_3}).
We keep image vectors satisfying $|n| \le n_{\rm max}$,
with $n_{\rm max}$ determined separately for each
sum and for each value of $ML$, using the following criterion.
For $|n|< n_{\rm max}$, 
a shell in image space of radius $|n|$ and unit
thickness should give a contribution larger
than the desired precision times the leading ($|n|=1$) term.
Thus we keep all $n$ satisfying
\begin{eqnarray}
  [4\pi |n|^2 ] \times \frac{K_1(|n| ML)}{|n|} 
  &\ge& \epsilon \times [6 K_1(ML)] 
  \ \text{ for } \delta^\textrm{FV}_1 \,,
  \label{eq:r}
  \\ {}
  [4\pi |n|^2 ] \times K_0(|n| ML)
  &\ge& \epsilon \times [6 K_0(ML)] 
  \ \text{ for } \delta^\textrm{FV}_3\,,
  \label{eq:r-3}
\end{eqnarray}
with $\epsilon = 1.0 \times 10^{-14}$ since we
want double-precision accuracy.
Note that on the left-hand side we are assuming that the
dominant contribution is from spatial images, which is a
reasonable approximation.

The values of $n_{\rm max}$ depend, of course, on $ML$.
For $\delta^{\rm FV}_1$ we find, for example, that for
$ML=0.4$, $2.0$ and $4.0$ that $n_{\rm max}=93$, $18$ and
$10$, respectively. The values for $\delta^{\rm FV}_3$ are
slightly larger, although comparable.
%\footnote{%
%
%Note that for the fits themselves, we need only values of
%$ML$ down to about $3.0$, while the smaller values 
%are used in the following plots to show how the FV corrections 
%behave at small $ML$.}
%

Using a single core of the Intel i7 920 CPU, it takes about two months
to run the analysis code on our full dataset
in double precision including FV corrections,
at a sustained speed of about 0.5 Gigaflop/s.
By contrast, using the Nvidia GTX 480 GPU, 
and optimizing our code using the CUDA library,
we obtain a sustained performance 
in double precision of 64.3 Gigaflop/s,
38\% of peak.

\begin{figure}[htbp]
  \includegraphics[width=17pc]
  {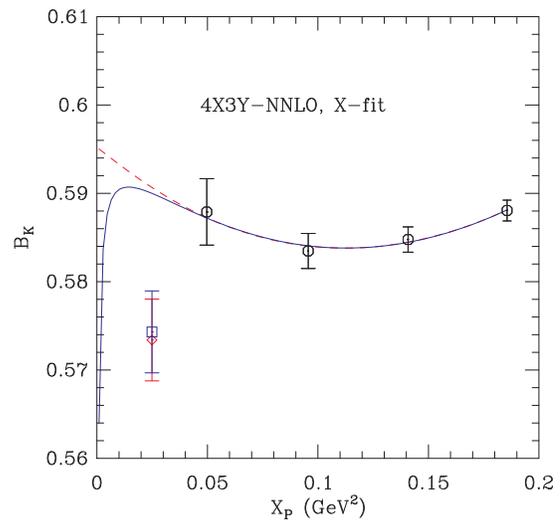}
  \caption{ $B_K(1/a)$ vs. $X_P$ on the C3 ensemble for $a m_y=0.05$.  
    The fit type is 4X3Y-NNLO. % in the SU(2) SChPT analysis.  
    The dashed [red]
    line shows the result of fitting to the infinite volume form,
    while the solid [blue] line shows the fit with FV corrections included. 
    The [red] diamond and [blue] square correspond 
    to the ``physical'' $B_K$ values obtained (as described in the text)
    from the infinite and finite volume fits respectively.
    }
  \label{fig:C3}
\end{figure}

We first compare the results of fitting with and without FV corrections
on the C3 ensemble. We note that on this ensemble (and on C3-2) we have
about 9 times the number of measurements as on the finer ensembles
and thus considerably smaller statistical errors.
Figure~\ref{fig:C3} shows the X-fits for our heaviest valence
strange quark ($a m_s^{\rm val}=0.05$). 
%We recall that $X_P$ is the
%squared mass of the pion composed of valence $d$ and $\bar d$ quarks.
Although our lightest valence pion mass corresponds to $ML=2.7$ on 
these lattices, the difference between the infinite
and finite volume fits is not significant until much smaller values of $ML$.
This is in apparent contradiction with rule-of-thumb that one
has large FV effects when $ML < 3$.
This conundrum is resolved by the presence of taste breaking.
Almost all the pions which appear in the loops
have non-Goldstone tastes, and thus are heavier than the Goldstone
pion by shifts of $O(a^2)$. These shifts push
$ML$ up to values considerably larger than $3$, except for
a small contribution from the Goldstone pion itself.
%[This contribution is small because of the factor 
%of $\tau^P=1/16$---see Eq.~(\ref{eq:q_1}).]

The net result is that fitting with the FV form leads to a
very small shift in the extrapolated ``physical'' values of $B_K$,
as shown in the first row of Table~\ref{tab:bk-fv}. 
These values are obtained by using the fit
function to (a) extrapolate to the physical valence $d$ mass,
(b) set all taste-splittings to zero,
(c) set the light sea-quark pion mass-squared
to its physical value, and
(d), in the case of the FV fit, set the volume to infinity.
We use quotes around ``physical'' because these values 
of $B_K$ have yet to be extrapolated to the physical strange quark 
mass and to the continuum limit.
Note also that $B_K$ is here matched to a continuum
operator renormalized at a scale $1/a$, so that results from
different lattice spacings are not directly comparable.

%---------------------------
% B_K: finite volume effect
%---------------------------
\begin{table}[htbp]
  \caption{``Physical'' $B_K(\text{NDR},1/a)$ 
obtained as described in the text from fits without and
with finite volume corrections, and the percentage difference
between the two.
%All fits use SU(2) SChPT and are of type 4X3Y-NNLO.
The valence strange quark mass is set to
to its heaviest value on each ensemble.
%(e.g. $am_y=0.05$ for ensemble C3).
Errors are statistical.
    \label{tab:bk-fv}}
\begin{ruledtabular}
\begin{tabular}{ l l l l}
ID    & $B_K$       & $B_K$(FV)   & $\Delta B_K$ \\
\hline
C3    & 0.5734(46)  & 0.5743(46)  & +0.159(2)\% \\
C3-2  & 0.5784(46)  & 0.5785(46)  & +0.0319(3)\% \\
F1    & 0.5225(111) & 0.5199(110) & -0.505(14)\% \\
S1    & 0.4914(65)  & 0.4898(65)  & -0.329(6)\% \\
U1    & 0.4780(92)  & 0.4757(92)  & -0.474(14)\% \\
\end{tabular}
\end{ruledtabular}
\end{table}

From the Table, 
we see that the FV fit leads to a result that is 0.16\% higher
on the C3 ensemble. Although this difference is much smaller
than the statistical errors in the individual results, 
it is statistically highly significant. This is possible
because of the high degree of correlation between the
infinite volume and FV fits. The same holds true on the other
ensembles.

\begin{figure}[htbp]
  \includegraphics[width=17pc]
  {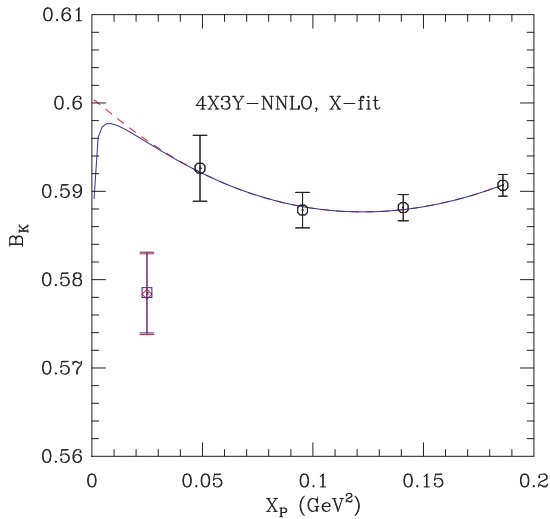}
  \caption{ $B_K(1/a)$ on the C3-2 ensemble. 
    Notation as in Fig.~\protect\ref{fig:C3}.  }
  \label{fig:C3-2}
\end{figure}

Figure~\ref{fig:C3-2} shows the corresponding fits on ensemble of
larger lattices, C3-2.  As expected, the difference between the
infinite and finite volume curves does not become significant until
smaller values of the pion masses than in Fig.~\ref{fig:C3}.  The
``physical'' $B_K$ values resulting from these two fits are
essentially the same.  This supports our claim in SWME-1 that we can
essentially treat the C3-2 ensemble as having infinite volume when
calculating $B_K$.

A natural question is whether the FV fit leads to better agreement
between the C3 and C3-2 ensembles.  The answer is a qualified yes.
The FV shift does bring the two results closer, although it removes
only 20\% of the difference.  It should be noted, however, that the
results on these two ensembles, which are statistically independent,
agree within 1-$\sigma$ already before the FV correction. Thus the
improvement is marginal.

\begin{figure}[htbp]
  \includegraphics[width=17pc]
  {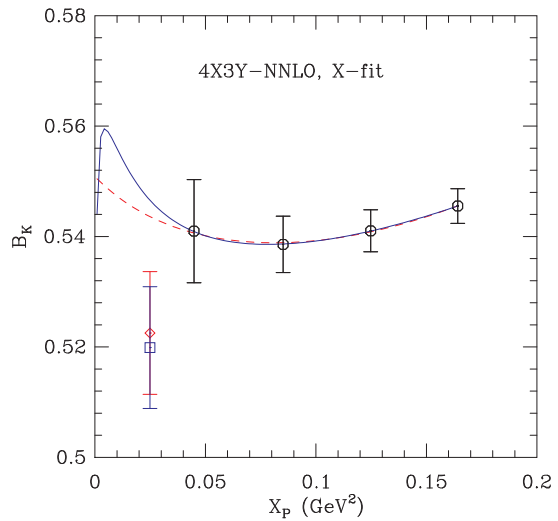}
  \caption{
    $B_K(1/a)$ on the F1 ensemble.
    Notation as in Fig.~\protect\ref{fig:C3}.
  }
  \label{fig:F1}
\end{figure}
\begin{figure}[htbp]
  \includegraphics[width=17pc]
  {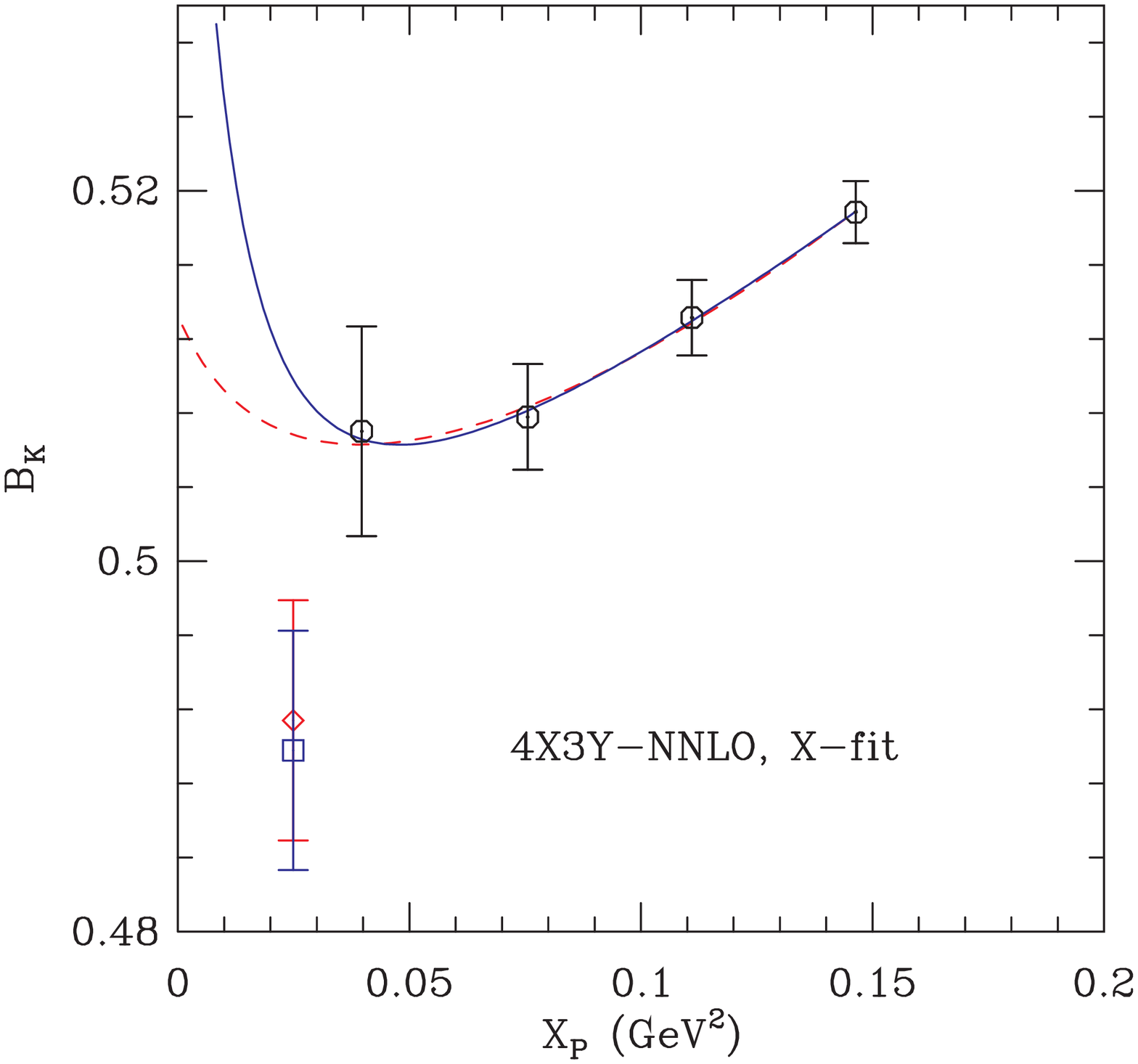}
  \caption{
    $B_K(1/a)$ on the S1 ensemble.
    Notation as in Fig.~\protect\ref{fig:C3}.
  }
  \label{fig:S1}
\end{figure}
\begin{figure}[htbp]
  \includegraphics[width=17pc]
  {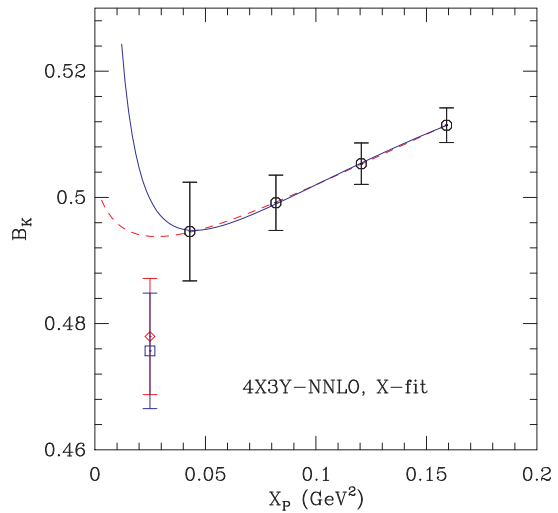}
  \caption{
    $B_K(1/a)$ on the U1 ensemble.
    Notation as in Fig.~\protect\ref{fig:C3}.
  }
  \label{fig:U1}
\end{figure}
%

%We now turn to the finer lattices. 
Figures~\ref{fig:F1},
\ref{fig:S1} and \ref{fig:U1} show the fits with and without
FV corrections on the F1, S1 and U1 ensembles, respectively.
We see that the FV corrections are of opposite sign to those on the C3
and C3-2 ensembles, that FV effects set in at a larger value of
$X_P$ as one reduces $a$, and that FV effects at the physical value of
$X_P$ grow rapidly as $a$ is reduced.
All these changes are due to the reduction in 
the size of taste breaking as $a$ is decreased,
so that the values of $ML$ for 
non-Goldstone pions are much closer to those of the Goldstone pion,
enhancing FV effects. 
%, which scales as $a^2\alpha_s(1/a)^2$.
%The reduction is by more than an order of magnitude between the coarse
%and ultrafine lattices~\cite{wlee-10-3}. 
It turns out that the enhancement is greater for terms contributing
positively than those contributing negatively, leading to the change
in sign.
It is clear from the curves that our lightest valence
pions are close to the smallest values that can be used without 
encountering large FV corrections.

The corresponding values of ``physical'' $B_K$ are given in Table
\ref{tab:bk-fv}. The FV shifts are now larger, as large as $0.5\%$ in
magnitude, although still small compared to other systematic errors.
We stress that one cannot gauge the size of the FV shift at the
physical value of $X_P$ by looking at the difference between the two
curves for this value. This is because the fits are being done at
larger values of $X_P$, where the FV corrections are smaller.

\begin{figure}[htbp]
  \includegraphics[width=17pc]
  {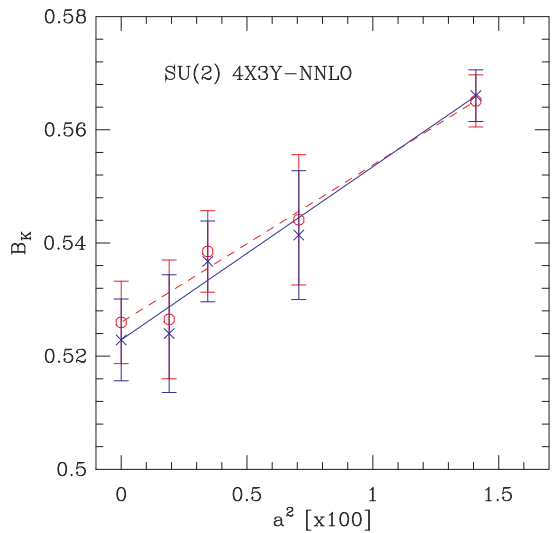}
  \caption{Continuum extrapolation of $B_K({\rm NDR},2\;{\rm GeV})$.
    [Red] octagons show results
     obtained using SU(2) 4X3Y-NNLO SChPT fits without the
    finite volume corrections, while [blue] crosses show the
    results obtained with FV fits.
    The points at $a=0$ are obtained by linear extrapolation in $a^2$.}
  \label{fig:scaling}
\end{figure}

In Fig.~\ref{fig:scaling}, we show how the continuum limit is impacted
by the inclusion of FV corrections. Here we plot $B_K$ after
extrapolation to the physical valence strange quark mass (using 3Y
fits), and after running in the continuum to a common scale of
$2\;$GeV. The overall effect of fitting with FV corrections is to
reduce the continuum result from $0.5260(73)$ to $0.5229(72)$, 
i.e. by $0.59 \pm 0.01\%$.
%
%0.587980973798095478 +/- 0.00764075255079602301 %.
%
%
%
%

%\section{Conclusion \label{sec:conclude}}

%
In conclusion, using GPUs we have been able to improve the
chiral extrapolation of our data for $B_K$ by the inclusion
of FV effects arising from pion loop diagrams.
The net result is that, after continuum extrapolation, we find that
including the FV shifts leads to an (infinite volume) result for
$B_K$ which is 0.59\% smaller than that obtained when fitting with
the infinite volume forms. The smallness of this shift confirms that
our values of $ML$ (which range down to $\approx 2.7$) are (barely)
large enough. In this regard, we are helped by the presence of
taste breaking, so that most of the pions appearing in the loops are
heavier than our lightest pion.

As is well known, one-loop ChPT forms can
underestimate the size of FV shifts, due to the contributions
from higher-order terms. This underestimate can be
by as much as a factor of 2 for our range of $ML$
(see, e.g., Ref.~\cite{Colangelo:2005gd}).
Thus, our new procedure for accounting for FV errors
is to take the central value from
the fits including FV effects, and take the systematic error
to be the size of the FV shift, namely 0.6\%.
This estimate is a little smaller than our previous
estimate of the FV error, 0.89\%, which was based on the difference
between the results on the C3 and C3-2 ensembles.
%\footnote{%
%
%  Previously we did not shift the central value of $B_K$, but rather
%  used that obtained fitting to infinite volume forms.  This was
%  because we did not know how the finite volume shift would depend
%  upon $a$.}
%
We think that our new procedure is more reliable, since it  
includes a continuum extrapolation, and is based on the theoretically
expected functional form (which, in particular, includes the
expected dependence on $a$). Our previous estimate also did not account
for the possibility that the difference on the two ensembles 
C3 and C3-2 could be statistical.

%In light of these considerations, we update our result for
%$B_K$, now based on four lattice spacings, to
%\[
%\hat{B}_K = B_K (\text{RGI}) = 
%0.7160 \pm 0.0099(\text{stat}) \pm 0.0345(\text{sys})
%\]
%with the error budget given in
%Table \ref{tab:err-budget}. 

%                    
%

\begin{acknowledgments}
C.~Jung is supported by the US DOE under contract DE-AC02-98CH10886.
The research of W.~Lee is supported by the Creative Research
Initiatives program (3348-20090015) of the NRF grant funded by the
Korean government (MEST).
The work of S.~Sharpe is supported in part by the US DOE grant
no. DE-FG02-96ER40956.
Computations for this work were carried out in part on QCDOC computers 
of the USQCD Collaboration at Brookhaven National Laboratory.
The USQCD Collaboration are funded by the Office of
Science of the U.S. Department of Energy.
\end{acknowledgments}

%----------
% Appendix
%----------
\appendix
%

%-----------
% reference
%-----------
\bibliographystyle{apsrev} %%% physical review
\bibliography{ref} %%% ref.bib file

\end{document}